\documentclass[twoside]{article}
\usepackage{fleqn,espcrc2}
\usepackage{epsfig}
\usepackage[figuresright]{rotating}

\input boxedeps.tex
\SetOzTeXEPSFSpecial
\SetepsfEPSFSpecial
\HideDisplacementBoxes
\newcommand{\etal}{{\em et al.}}

\newcommand{\gev}{\hbox{ GeV}}

\newcommand{\tev}{\hbox{ TeV}}

\newcommand{\km}{\hbox{ km}}

\newcommand{\eqn}[1]{(\ref{#1})}

\newcommand{\pl}[3]{{\em Phys. Lett.\/} {\bf #1} (19#3) #2}

\newcommand{\pr}[3]{{\em Phys. Rev. D\/}{\bf #1} (19#3) #2}

\newcommand{\np}[3]{{\em Nucl. Phys.\/} {\bf #1} (19#3) #2}
\newcommand{\npbps}[3]{{\em Nucl. Phys. B (Proc. Supp.)\/} {\bf #1} (19#3) #2}

\newcommand{\zp}[3]{{\em Z.~Phys. C\/}{\bf#1} (19#3) #2}

\newcommand{\astropp}[3]{{\em Astropart. Phys.\/} {\bf #1} (19#3) #2}
\newcommand{\ib}[3]{{\em ibid.\/} {\bf #1} (19#3) #2}
\newcommand{\iauc}[4]{{\em IAU Circular\/} #1 (\ifcase#2\or January\or
February\or March\or April\or May\or
June\or July\or August\or September\or October\or November\or
December\fi\ #3, 19#4)}

\newcommand{\yadfiz}[4]{{\em Yad. Fiz.\/} {\bf #1} (19#3) #2 [English
transl.: {\em Sov. J. Nucl. Phys.\/} {\bf #1} (19#3) #4]}
\newcommand{\jetp}[6]{{\em Zh. Eksp. Teor. Fiz.\/} {\bf #1} (19#3) #2
[English translation:
         {\it Sov. Phys.--JETP } {\bf #4} (19#6) #5]}

\newcommand{\hepph}[1]{(electronic archive: hep--ph/#1)}
\newcommand{\astro}[1]{(electronic archive: astro--ph/#1)}
\relax
	\newcommand{\Qpr}[3]{{Phys. Rev. D.} {\bf #1,} #2 (19#3)}									 %
	\newcommand{\Qpl}[3]{{Phys. Lett.} {\bf #1,} #2 (19#3)}						 %
	\newcommand{\Qprl}[3]{Phys. Rev. Lett. {\bf #1,} #2 (19#3)}				 %

\begin{document}

\title{ Ultra-High Energy Neutrinos: A Review of Theoretical and Phenomenological Issues   }
\author{Raj Gandhi\address{Harish-Chandra Research Institute, \\
Jhusi, Allahabad, India 211019.}}%


\begin{abstract}

We review the phenomenology of UHE neutrino detection.  The  motivations for looking for such neutrinos,
 stemming from observational evidence and from the potential for new physics discoveries are enumerated,
 and their expected sources and   fluxes are given. Cross-sections with nucleons at energies all the way upto $10^{20}$ eV and the attenuation of fluxes in the
Earth, both of which are physics issues important to their detection, are discussed. Finally, sample event-rates for extant and planned  
Water/Ice Cerenkov detectors are provided. 

\end{abstract}

\maketitle


\section{Introduction}
It is now widely beleived that the recent Super Kamiokande (SK) result {\cite{fuk,sobel}} of an anomaly in the flavour ratios and zenith angle dependance of the
atmospheric neutrino flux provides the firmest signal yet of physics beyond the Standard Model (SM)\footnote{The first hints of this anomaly were
provided by atmospheric neutrino data from the IMB detector, \cite{IMB}} . The significance of this can be gauged by the fact that  
a signal for such physics
 has been sought for in all the varied and intensive experimental probes that the many sectors  of the SM have been subjected to over more than  twenty-five years. That this signal comes from the neutrino sector of the theory  perhaps  assumes increased significance when considered in conjuntion 
with the fact that two other existing experimental anomalies which persist, 
the solar  deficit \cite{solar} and the LSND result \cite{lsnd} are also 
within the neutrino sector.

  The detailed interpretation of these anomalies
and their relation with each other  is still
a matter of intense  and ongoing theoretical and experimental activity. However,  neutrino masses,
mixings and consequent oscillations have provided the simplest framework for understanding the experimental results \cite{smir}.  What may be said with a reasonable degree of conviction is that once their interpretation is clear, a hitherto unprecedented window into the nature of the physical theory beyond the SM will probably have been opened.

It is within this context, perhaps, that one should view the theoretical and experimental efforts that are ongoing in the area of ultra high energy (UHE) neutrino
physics (for reviews see \cite {gqrs,h1,r98}). If the physics excitement that we have uncovered in our study of low-energy neutrinos is an indicator, then the study of UHE neutinos should pay rich dividends.

 In view of the fact that data collection
and further upgradation is ongoing for the first generation of UHE neutrino detectors ( AMANDA \cite{aman} and  BAIKAL \cite{bai}) and that planning, design, testing and deployment
is underway for some others (AUGER \cite {aug}, NESTOR \cite{nest}, ICECUBE \cite{ice},  ANTARES \cite{ant},   RICE \cite{rice}), and NEMO \cite{nemo})
 definitive results of these efforts are still in the future. However, besides the other  specific reasons from astrophysics and cosmic ray physics (some of which we discuss below) 
which provide motivation for such experiments, we note that the energy range covered by these experiments ($E_{\nu} \simeq 1$ Tev to $10^9$ TeV or higher) 
offers an unprecedented opportunity for particle physics at energies significantly beyond the scope of terrestial accelerators. Although progress is expected to be slow, the potential for serendipitous discovery is undoubtedly high.

In what follows, we  give specific reasons why the search for UHE neutrinos 
can be expected to yeild positive results. Possible sources and their fluxes
are discussed.
 Salient features pertinent to 
the phenomenology of  detection are then outlined, and sample rates provided.
 
\section{Why should we search for and expect to detect UHE neutrinos?}

In addition to the general particle physics motivations mentioned above, there exist  specific reasons spanning several fields of research (astrophysics, astronomy,  cosmic-ray physics and particle physics) for pursuing the search for UHE neutrinos. Motivation for their detection comes both from observational evidence hinting  towards  their production in astrophysical sites and the potential for new physics discoveries. 

\subsection{The Observed Cosmic-Ray Spectrum:}

 Perhaps one  the strongest reasons comes from the observed cosmic-ray (CR)
spectrum. We briefly discuss its observational features prior to making the connection to UHE neutrinos.
 
 Over a period of several decades, a sizeable and impressive body of observations spanning twelve orders of magnitude in energy have been compiled by workers in this field. (For  general discussions see \cite{ber,gai}, and for updates on observational efforts and data see \cite{icrc99}.)
An examination of the all-particle 
CR spectrum   reveals a  power-law behaviour  over almost the entire spectral range, with breaks at what is referred to as the ``knee'' (corresponding to $E_{primary} \simeq 4 \times 10^{15}
$ eV) and the ``ankle''(corresponding to $E_{primary} \simeq 10^{19}
$ eV).

 Upto the knee region, the spectrum exhibits a power-law with index
$-2.7$, there onwards steepening to about $- 3$, and flattening again in the ankle region. The steep fall in the flux and the consequent increasing  difficulty in detection  is reflected in the fact that the number of primaries falls from about one particle per m$^{2}$ per sec at 
energies of  about 10$^{11}$ eV to roughly one per km$^2$ per century 
around the ankle (10$^{19}$ eV). 

It is thus necessary to employ very different 
techniques of detection depending on the energy of the primary. The direct 
observation of a CR primary is only possible in a detector mounted on a spacecraft
or a balloon, due to interactions in the atmosphere. 
 Such detection, however
can collect enough statistics  only upto primary energies of about 10$^{14}$ eV. At higher energies, indirect methods involving detection of secondaries produced by interaction of the primaries in the atmosphere is necessary. The most widely used indirect method has been the deployment of Extensive  Air Shower (EAS)
arrays (for example  Yakutsk \cite{yak6} and AGASA \cite {aga8}). These sample a lateral cross-section of the multi-particle shower initiated by a CR primary high up in the atmosphere.

 Clearly, the determination of the energy of the primary from the charged hadrons, electrons, muons and photons detected by the
ground-based array is a non-trivial task. However, present Monte Carlo 
techniques allow this to be accomplished with an accuracy of about $30\%$.
Still less trivial is the determination of the nature of the primary, which is 
complicated by fluctuations from shower to shower. In recent years another
experimental technique, which focusses on the detection of secondary nitrogen fluorescence 
radiation excited by shower secondaries has been  succesfully implemented by Fly's Eye \cite {fly7}. 

Using the modes of detection discussed above, a general but incomplete picture
of the nature of CR primaries has emerged. Essentially, below primary energies of $10^{14}$ eV, the composition is fairly accurately known, from a variety of direct detection experiments, to be $98\%$ hadrons, mainly protons. Above these
energies, where only indirect detection techniques can be employed,
the composition appears to retain a significant hadronic fraction, even though the question as to whether the primaries are protons or heavier nuclei is yet unresolved.  
Very low statistics hampers  any definitive conclusions about the composition of CR beyond the knee upto  
the highest energies (at and above the GKZ cutoff, which is discussed below). However, we stress that 
all or most of the by now numerous EAS events appear to have a muon content consistent with hadronic
primaries.

For our purpose it is sufficient to say that it appears almost 
certain  from the CR observations  that 
there are astophysical sites in the universe which acclerate hadrons upto
  energies of $10^{15}$ eV, and there is a reasonable probability that they are accelerated to 
higher energies, perhaps all the way upto $\approx 10^{20}$ eV. The mechanisms by which this happens are not completely understood.  However, hadronic collisions, like $p + p$ or $p + \gamma$,  always  result in 
copious pion production. These follow the decay chain $\pi \rightarrow \mu +
\nu_{\mu}, \mu \rightarrow e + \nu_{\mu} + \nu_e$ resulting in a flux ratio
of 2:1 for $\nu_{\mu} : \nu_{e}$. The neutrino is  expected to retain about 20$\%$ of the energy of the parent pion on the average, and  in general, UHE neutrino fluxes are  guaranteed if hadronic collisions play a role in UHE CR
production. The precise shape of  the spectrum, of course, depends on the nature of the source and the process of acceleration.

\subsection{ The GKZ Effect}

In addition to neutrino production at the source,
a UHE CR  proton with energy in excess of about $5 \times 10^{19}$ eV traversing 
interstellar space will interact with the Cosmic Microwave Background (CMB)
to photo-produce pions, and hence neutrinos. (Representative fluxes
are given in \cite{proth}.) The interaction in question, 
the $\Delta$ resonance for single and multiple pion production, sets in at a 
center-of-mass energy of about $1.5$ GeV and  is responsible for an expected abrupt fall-off in the proton primary spectrum at and above these energies. This was first pointed out by Greisen, Kuzmin and Zatsepin \cite{gkz21,gkz22} and is called the GKZ cutoff. 
\footnote{ This cutoff is at present associated with one of the most significant puzzles in CR physics (see \cite{gsigl} for a recent review), because CR events with energies in excess of 
$10^{20}$ eV have been convincingly detected by several experiments employing different detection techniques.  Although we do not go into this puzzle in any detail here, or the apparent absence of a GKZ cutoff signalled by the highest energy CR, we  note  that the existence of the cutoff is demanded by 
well-tested low-energy SM physics. Additionally, the observation of such
post-GKZ events has relevance  for UHE neutrino-nucleon cross-sections in
non-standard scenarios, which we discuss briefly in Section 5. }

\subsection{ Galactic CR Interactions with Inter-stellar matter}

 A  diffuse UHE neutrino flux is also expected from interactions of galactic CR
 with interstellar matter, at energies around and below the knee region. For fluxes
and related references, see \cite{proth} 

\subsection {Gamma-Ray Bursts}

Yet another potentially important source of UHE neutrinos could be Gamma-Ray
Bursts (GRB) \cite{eli}. There appears to be observational support for a ``fireball''
model for GRBs \cite{pij364}. The physics of this involves an initial merger or collapse of blackholes, neutron stars or some other highly magnetized
compact object. The collapse to a small radius object is followed by a very rapid expansion, by about a factor of $10^5$ in a time frame of the order of a 
second. This ultra-relativistic acceleration of the plasma of protons, electrons, positrons and photons leads, at some radius at which the plasma is optically thin to radiation which is detected as the GRB. At the same time, second-order
Fermi shocks cause charged particle acceleration, and the ensuing $p-\gamma$
interactions lead to neutrino fluxes, which have been estimated in \cite{waxb,pij366} for instance (See figure \ref{fig:fluxes}).

\subsection{ Active Galactic Nuclei}

Active Galactic Nuclei (AGNs) are also expected to be an important source of 
UHE neutrinos. These are a class of highly compact bright objects powered
presumably by black holes causing acceleration and accretion of matter, 
characterised usually by high powered jet emmission. High energy gamma-rays
(MeV, GeV and TeV) are expected and have been observed from $\approx 40$
AGN sites. The acceleration of hadronic matter, as in proton-blazar models, again leads to subsequent 
$p-\gamma$  (and perhaps $p-p$) interactions, from which neutrinos are expected. (Figure \ref{fig:fluxes}). 
We note, however, that if only electrons are accelerated, as in the electron-blazar models, then a neutrino flux will be absent.

\subsection{ Topological Defects and Decays of other Massive Relics}

There also  exist mechanisms which do not involve the acceleration of matter but can still yeild UHE fluxes of protons, photons and neutrinos. These are the ``top-down'' (TD) scenarios (for a comprehensive review of these see \cite{pij}) in which CR beyond the knee could originate
 in the decay of GUT scale massive particles produced due to topological defects like monopoles, cosmic strings, etc. Massive unstable relic particles \cite{b2}
originating in various unification scenarios like string and supergravity theories which give rise  to UHE neutrino fluxes have received attention recently in connection with the observed absence of a GKZ cut-off see also,  for instance  \cite{subir,gel}. In addition, the photon fluxes from such exotic decays may be energetic enough to 
pair-produce muons off the CMBR photons \cite{ak}, with consequent neutrino production. Although such exotic sources produce low fluxes, it may be possible to see a signal in future detectors; see for instance the discussion in \cite{almun} regarding  their detectability in ICECUBE.     
 Some  representative fluxes for such sources  are shown in
Figure \ref{fig:fluxes}.

\subsection{Additional Physics Motivations}
The detection of UHE neutrinos from all the above sources is expected to 
help answer many important questions in CR physics and the astrophysics of 
highly energetic sources.  Their detection above $10^{16}-10^{18}$ eV
from AGNs would provide important support for hadronic blazar (versus electron
blazar) models. 

 Secondly,
the observation of an isotropic  neutrino flux beyond GKZ energies
would signal that the events are due to universal CR activity rather than a
nearby single source and would shed light on the nature of the primaries 
responsible for these events.

 In addition, as emphasized in \cite{waxb},
very sensitive tests of gravitational couplings are possible via the detection of neutrinos co-related with GRB's. 

Finally, important confirmation of the 
SK atmospheric result, which now firmly indicates that muon-neutrinos are most likely oscillating to tau-neutinos, is possible.  UHE Neutrinos which are produced by any of the modes or sources discussed above are not expected to have any significant component of $\nu_{\tau}$, originating as they do in $p-p$ and/or
$p-\gamma$ interactions. However, the ultra-high energies
and mega-parsec distances they travel prior to detection, when folded in
with SK values for mass-squared differences and mixing angles, lead to 
a $\nu_{\tau}$ component in the flux which is as large as the $\nu_{\mu}$ 
and the $\nu_e$ components. It may be possible to detect this via the 
observation of the tau lepton it can produce in the detector \cite{pak}. In addition,  the shadowing (in the earth)  for $\nu_{\tau}$ is interestingly 
different compared to that for $\nu_{\mu}$ 
and  $\nu_e$, as pointed out in \cite{halzen}. Specifically,  the $\nu_{\tau}$
component of the flux does not get significantly absorbed in the rock, but its 
energy spectrum gets modified, and its zenith angle distribution is relatively flat. The detection possibilities for such UHE  $\nu_{\tau}$ are discussed in 
\cite{sar,ath}.

\section{UHE Neutrino Fluxes}

\begin{figure}[tbp]
\centerline{\BoxedEPSF{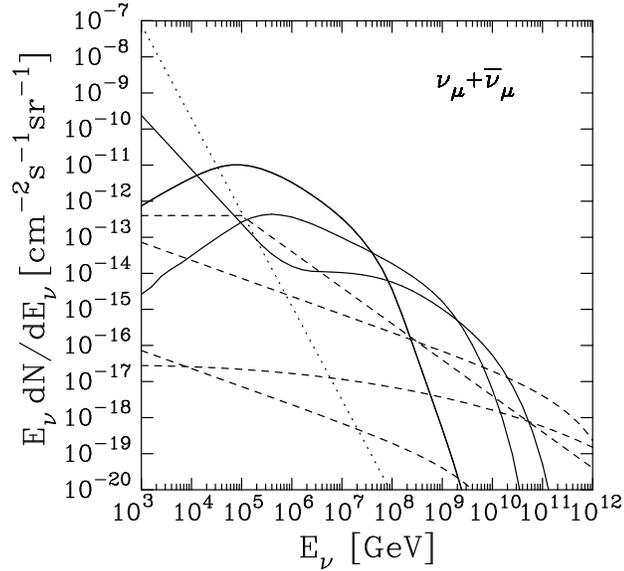  scaled 450}}
\caption{Muon neutrino plus antineutrino fluxes scaled by neutrino 
energy at the Earth's surface.  Solid lines represent AGN fluxes.  In 
decreasing magnitude at $E_\nu=10^3$ GeV, they are AGN-M95, AGN-SS91 
scaled by 0.3, and AGN-P96 ($p\gamma$).  The dashed lines, in the same 
order, represent the GRB-WB, TD-WMB12, TD-WMB16, and TD-SLSC fluxes.  
The dotted line is the angle-averaged atmospheric (ATM) neutrino 
flux.}
\label{fig:fluxes}
\end{figure}

In Figure \ref{fig:fluxes} we show some of the flux calculations for UHE neutrinos 
 from AGNs, GRBs and TD sources. We refer the reader to the references
for details on the models used to obtain the predicted fluxes.  The fluxes are labeled as AGN-M95 \cite{r49},
AGN-SS91 \cite{r56}, AGN-P96 \cite{r48} for Active Galactic Nuclei, GRB-WB
\cite{r50} for Gamma Ray Bursts and TD-WMB12 \cite{r52}, TD-WMB16 \cite{r52},
TD-SLSC \cite{slsc} for top-down models. References and spectra for the remaining
sources  mentioned above may be found in \cite{proth}, along with a discussion of their origins and sources.

It is important to stress here that fluxes for all the three sources 
in Figure \ref{fig:fluxes} are still being modified as our understanding of the nature of the source, the interactions involved and the constraints placed by observations on them improve. For instance, it has been pointed out \cite{wb} that observations of the CR  spectrum place important constraints on the neutrino flux from 
sources which are optically thin with respect to neutrons and photons. This argument could significantly restrict several  AGN models, but the exact upper bound is
presently being debated. In \cite{mpr}, for instance, it is argued that this
bound depends on the assumption that the overall CR injection spectrum
$dN/dE \propto E^{-2}$ upto  energies of $10^{19}$ eV, and that injection spectra for many sources of CR producing shocks may be different, and, for a given source,
may not extend all the way upto $10^{19}$ eV. The importance of using CR and
other observations to constrain the physics occuring in the sources, and the 
neutrino fluxes emmitted as emphasized in \cite{wb}  however, cannot be disputed.

We also refer the reader to the discussion of fluxes from 
TD models and decaying massive relics in \cite{subir2}, where problems with 
some of the existing  flux calculations are pointed out.

\section{The Detection of UHE Neutrinos: General Considerations}

Having discussed potential sources for UHE neutrinos, we next review
some salient points relevant to their detection.

 The main mode of detection discussed in the literature and implemented at extant detectors has been the observation of long range muons produced via charged current (CC) neutrino-nucleon
interactions. The low fluxes necessitate the deployment of large volume 
water or ice detectors like AMANDA \cite{aman}, BAIKAL \cite{bai}, NESTOR 
\cite{nest} and ANTARES \cite{ant}. 

These are shielded from above by several 
kilometers of water-equivalent (kmwe) rock volume in order to supress the otherwise overwhelming background from muons produced by CR interactions in the atmosphere. Indeed, for a considerable portion of the UHE range (100 TeV and lower), the detection of downward moving muons by Cerenkov radiation in water produced
by contained events is less advantageous than that of upward moving ones produced in the rock below the detector. The emphasis in existing detectors has thus been on
looking mainly for muons which, after losses due to passage in the earth, still
retain sufficient energy to be observed above threshold. In general, a 10 TeV
muon will travel several km in rock before its energy is degraded down to 
1 TeV, which is a typical detector threshold. \footnote{A charged particle moving through rock suffers two types of energy losses, continuous and discrete. Continous losses are mainly due to ionization of the medium, and their rate depends weakly (logarithmically) on the energy. Discrete losses stem from bremstrahllung, electromagnetic interactions with nuclei and direct $e^+e^-$ production.
In general, such losses are proportional to the energy, but for muons gain
importance only at higher energies.  The most important loss mechanism, 
bremstrahllung, is proportional to the inverse square of the mass of the 
charged particle, and hence is highly suppressed for muons relative to electrons.}

 Whereas serious neutrino astronomy
results are not expected with present detector volumes,  the actively  underway  upgradation  of existing facilities
and the addition of new ones should yeild significant physics results
in the next few years.   The first AMANDA \cite{aman} detector has seen about 170 UHE
atmospheric neutrino events, and  its upgraded version, AMANDA II is presently taking data. A km$^{3}$ extension, ICECUBE \cite{ice}, is also planned. BAIKAL
\cite{bai} has also observed UHE neutrinos and set limits on the UHE fluxes using its prelimnary observation, with analysis of another 3 years of data currently underway. The ANTARES project \cite{ant}, off the French coast, is exploring the design and implementation of 
a km$^3$ scale deep sea detector for underground muons. Similarly, NESTOR \cite{nest}
is doing the same southwest of the coast of Greece, while
NEMO \cite{nemo} is yet another planned deep sea detector off the southern Italian coast.

 The RICE project in the Antacrtic \cite{rice} will be a pioneering attempt to detect $\nu_e$ via radio waves. An UHE electron produced subsequent to a CC interaction in the rock transfers most of its energy to an electromagnetic shower. Positrons produced in the shower annhilate
and additional atomic electrons scatter into the shower, causing a charge imbalance which
corresponds to a ball of negative charge moving through the rock. The consequent Cerenkov emission gives radio waves ($\lambda \simeq 10$ cm), which are 
detected by receivers buried in the ice.

 Finally, the Pierre Auger Observatory
\cite{aug},
primarily designed to detect CR showers, should  also be able to detect UHE neutrino induced showers close to the horizon. Rate calculations for this array are given in \cite{r98}.

\begin{figure}[htb]
	\centerline{\BoxedEPSF{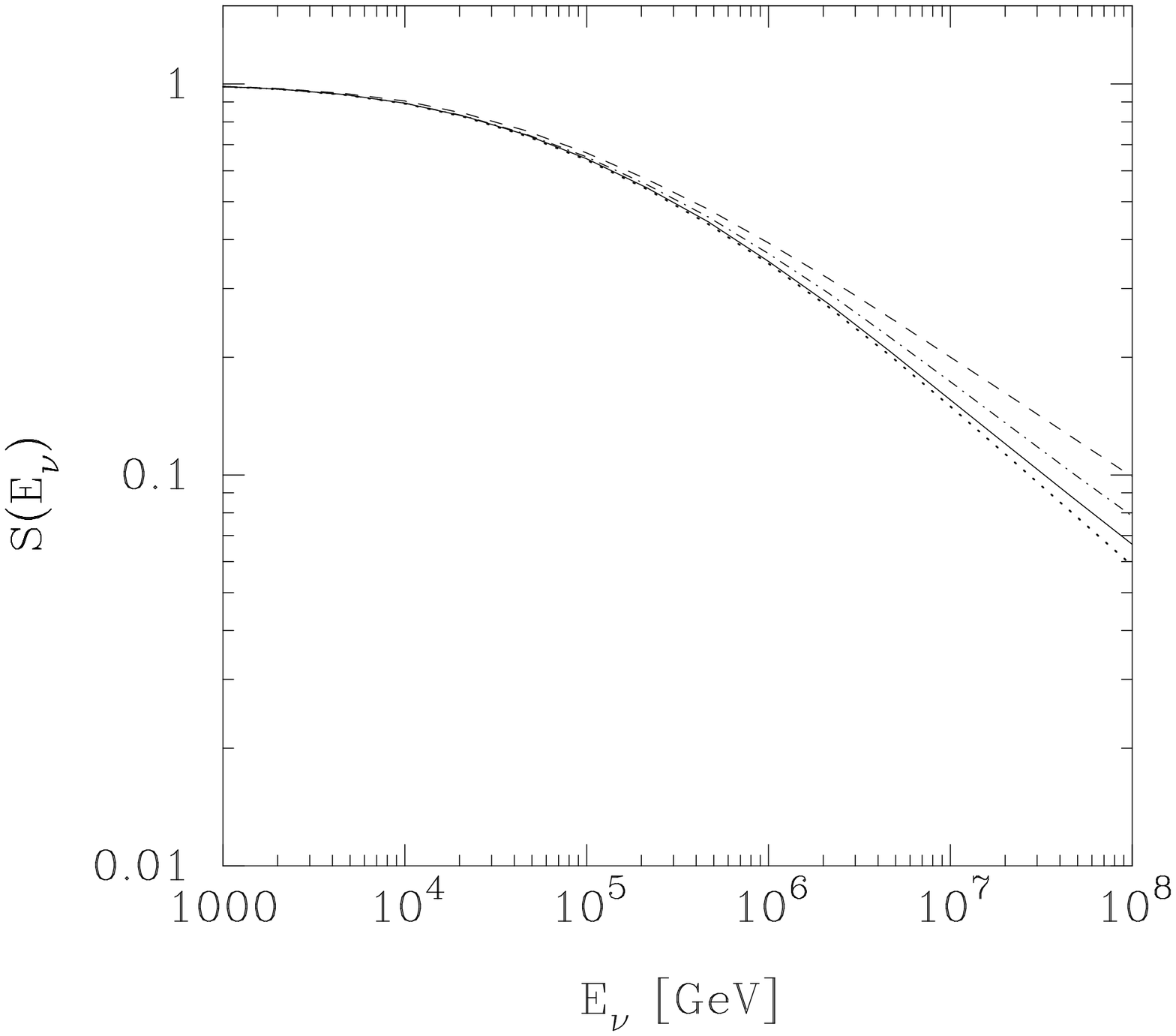  scaled 400}}
	\caption[shadow]{The shadow factor $S(E_\nu)$ for upward-going neutrinos
assuming that $\sigma=\sigma_{\mathrm{tot}}$ in \eqn{Sdef} for CTEQ-DIS (solid
line),
CTEQ-DLA (dot-dashed) and D\_ (dotted) parton distribution functions. Also
shown is the shadow factor using the EHLQ cross sections (dashed line).}
	\protect\label{fig:shadow}
\end{figure}

An important consideration for calculating event rates for muon
 neutrinos is the attenuation of neutrinos in the earth due to the 
rapid rise in the CC cross-section (discussed
below) with energy. The interaction length of a neutrino in rock
 is approximately equal to the diameter of the earth at $40$ TeV.
 A convienient quantity 
relevant to flux attenuation and event rate calculations is the 
'shadow factor',
$S$, which is defined to be an effective solid angle divided 
by $2\pi$ for upward muons and is a function of the energy-dependant 
cross-section for neutrinos in the earth \cite{gqrs}:
\begin{equation}
S(E_\nu)={1\over 2\pi}\int_{-1}^{\:0} d\cos\theta\int d\phi
\exp \left[-z(\theta)/{\mathcal L}_{{\mathrm int}}(E_\nu) \right].
\label{Sdef}
\end{equation} 
Here $ {\mathcal L}$ is the interaction length for the neutrino, defined by 
\begin{equation}
	{\mathcal L}_{\mathrm{int}}= \frac{1}{\sigma_{\nu 
	N}(E_{\nu})N_{\mathrm{A}}}\;,
	\label{Lint}
\end{equation}
where $N_{\mathrm{A}} = 6.022 \times 10^{23}\hbox{ mol}^{-1}=6.022 
\times 10^{23}\hbox{ cm}^{-3}$ (water equivalent) is Avogadro's 
number. $z$ represents the column-depth as a function of the nadir angle $\theta$.
Figure \ref{fig:shadow} (from Ref \cite{gqrs}) shows that from almost no attenuation
at $10^{12}$ eV, about $93\%$ of the flux is shadowed out at the highest energies at which CR have been observed ($\approx 10^{21}$ eV). It is this effect that makes the (low) downward rate for muons produced within the instrumented 
volume competetive with the upward rate at energies above $10^{15}$ eV, where the atmospheric background is essentially absent.

\section{The Neutrino-Nucleon Deep Inelastic Scattering Cross-Section at UHE}

Of crucial importance to  the attenuation and the event-rate calculations
is the UHE neutrino-nucleon DIS cross-section. This is given by

\begin{equation}
	\nu_\mu N \rightarrow \mu^- + \rm{anything}  ,
\end{equation}
where $N\equiv\displaystyle{\frac{n+p}{2}}$ is an
isoscalar nucleon,
in the renormalization group-improved parton model. The differential cross
section is written in terms of the Bjorken scaling variables
$x = Q^2/2M\nu$  and $y = \nu/E_\nu$ as
\begin{eqnarray}
	\frac{d^2\sigma}{dxdy} & = &  \frac{2 G_F^2 ME_\nu}{\pi} \left(
\frac{M_W^2}{Q^2 + M_W^2} \right)^{\!2} \nonumber \\
& &  \left[xq(x,Q^2) + x
\overline{q}(x,Q^2)(1-y)^2 \right] , \label{eqn:sigsig}
\end{eqnarray} where $-Q^2$ is the invariant momentum transfer between
the incident
neutrino and outgoing muon, $\nu = E_\nu - E_\mu$ is the energy loss in
the lab (target) frame, $M$ and $M_W$ are the nucleon and
intermediate-boson masses, and $G_F = 1.16632 \times 10^{-5}~\rm{GeV}^{-2}$ is
the Fermi
constant. The quark distribution functions are
\begin{eqnarray}
q(x,Q^2) & = & \frac{u_v(x,Q^2)+d_v(x,Q^2)}{2} + \nonumber \\
& &\frac{u_s(x,Q^2)+d_s(x,Q^2)}{2} +  \nonumber\\ & &  s_s(x,Q^2) + b_s(x,Q^2)
 \\[12pt]
	\overline{q}(x,Q^2) & = & \frac{u_s(x,Q^2)+d_s(x,Q^2)}{2} + \nonumber \\
& &   c_s(x,Q^2) +	t_s(x,Q^2),\nonumber
\end{eqnarray}
where the subscripts $v$ and $s$ label valence and sea contributions, and
$u$, $d$, $c$, $s$, $t$, $b$ denote the distributions for various quark
flavors in a  proton. At the energies of interest here, perturbative QCD corrections are small and can safely be neglected. A parallel calculation similarly leads to the neutral-current cross section.

In our calculations we have used  results from the $ep$ collider HERA
\cite{r9545, r9546,r9548,r9549} which have greatly enhanced our knowledge
of parton distributions and  are particularly significant for   the
present calculation.  
The usual procedure is to begin with parametrizations of parton distribution
functions obtained from data at low values of $Q^{2}$ and evolve them to
the desired high scale using the Altarelli-Parisi equations \cite{ap}.  For  UHE neutrino-nucleon
scattering, however,   the $W$-boson propagator forces increasing contributions from smaller and
smaller values of $x$ as the neutrino energy $E_{\nu}$ increases.  In
the UHE domain, the most important contributions to the $\nu N$ cross
section come from $x \sim M_{W}^{2}/2ME_{\nu}$.  Up to
$E_{\nu}\approx 10^{5}\gev$, the parton distributions are sampled
 at values of $x$ where they have been constrained by experiment.
At  higher energies, we require parton distributions at such
small values of $x$ that direct experimental constraints are not
available, not even at low values of $Q^{2}$.

Thus the  theoretical uncertainties that enter the evaluation of the UHE
neutrino-nucleon cross section arise from the low-$Q^{2}$
parametrization, the evolution of the parton distribution functions to
large values of $Q^{2}\sim M_{W}^{2}$, and the extrapolation to small
values of $x$. Of these, the last named contributes the greatest uncertainty.

In addition to the   traditional approach, followed, for instance,  in the CTEQ
\cite{cteq}
distributions, ( {\it i.e.}, to determine
parton densities for $Q^{2}>Q_0^{2}$ by solving the
next-to-leading-order Altarelli-Parisi \cite{ap}
evolution equations numerically) a second approach to small-$x$ evolution attempts to solve the
Balitski\u{\i}-Fadin-Kuraev-Lipatov
(BFKL) equation \cite{bfkl}, which is  a
leading $\alpha_s\ln(1/x)$
resummation of soft gluon emissions.    The BFKL approach predicts a
singular behavior in $x$ and a rapid $Q^{2}$-variation,
\begin{equation}
xq_s(x,Q^2) \sim \sqrt{Q^2} \:x^{-0.5}. \label{eqn:bfkl}
\end{equation}
On the other hand, applying the Altarelli-Parisi equations to singular input
distributions $\propto x^{-\frac{1}{2}}$ leads to a less rapid growth with $Q^{2}$,

\begin{equation}
xq_s(x,Q^2) \sim \ln{(Q^2)} x^{-0.5}. \label{eqn:apev}
\end{equation}

In the case of ultrahigh-energy neutrino-nucleon interactions, the region
of interest is small-$x$ and large-$Q^2$,
which requires a resummation of both
$\ln 1/x$ and $\ln Q^2/Q_0^2$ contributions.
An attempt to incorporate elements of the BFKL approach into the standard 
AP approach has recently been made in \cite{kms}. That this leads to cross-sections which are comparable (to within 10-15 \% at the highest energies)  with those obtained by next-to-leading order
AP evolution is an indication that barring radically new physics, the cross-section calculation even at the highest energies appears to be reasonably reliable.

\begin{figure}[htb]
		\centerline{\BoxedEPSF{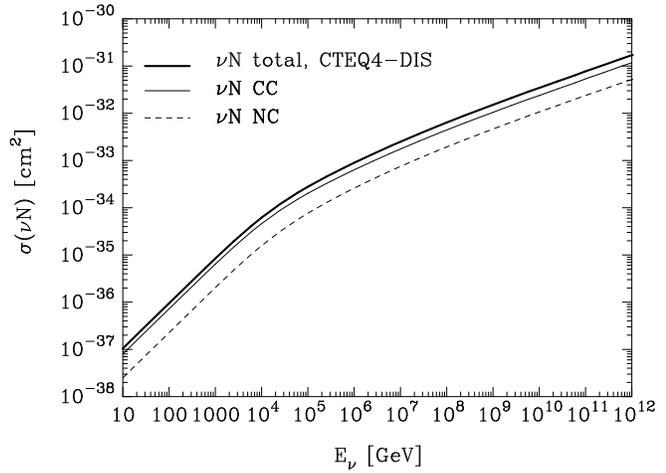  scaled 725}}
\caption{Cross sections for $\nu_{\ell} N$ interactions at high 
energies, according to the CTEQ4--DIS parton distributions: dashed 
line, $\sigma(\nu_{\ell} N \rightarrow \nu_{\ell}+\hbox{anything})$; 
thin line, $\sigma(\nu_{\ell} N \rightarrow 
\ell^{-}+\hbox{anything})$; thick line, total (charged-current plus 
neutral-current) cross section.}
	\label{fig:signuN4}
\end{figure}

 In Figure \ref{fig:signuN4} (from  \cite{r98}) we show the  
CC, NC and total cross-sections for $\nu - N$ interactions resulting from
the SM using the AP approach to small $x$ extrapolation.

Modifications to the cross-sections can occur, of course, as a result of
 physics beyond the SM. Many of these modifications are strongly constrained by
unitarity and for a broad class of plausible extensions are much smaller 
than the SM cross-section, as shown in \cite{r47}. However, since it appears certain that neutrino primaries with SM interactions cannot account for the ``post-GKZ''
 CR events mentioned earlier \footnote{For an interesting proposal that 
attempts to resolve this puzzle using neutrinos with SM interactions which
travel most, but not all of the distance to the earth before producing a hadronic primary  see \cite{tweil}} the possibility of a neutrino-nucleon cross-section rendered high
due to non-standard interactions has garnered substantial interest. This is in no small measure due to the fact that among the known particle candidates, the 
neutrino is the only possible primary that can make it without absorption
over the tens or hundreds of Mpc distances that such primaries  have almost certainly 
travelled prior to detection.
 Possibilities leading to 
 higher neutrino-nucleon cross-sections which have been discussed in the literature
and some of the mechanisms considered include leptoquark excitations \cite{r45}, superpartner contributions \cite{r42} and strong FCNC interactions \cite{r46}. One suggestion that has attracted a lot of interest recently is the speculation
that the string scale may be widely separated from the Planck scale, and be as low as a few tens of TeV \cite{add}. For neutrino-nucleon scattering, this 
leads to  a large (strong interaction-like) cross-section either due to higher
dimensional gravitational contributions \cite{jmr} or the possibility that
the post-GKZ events are already in the string regime, with an exponentially 
growing level density of Kaluza-Klein excitations \cite{dom,dom2,dom3}. For a
 useful general  discussion of  issues related to theories with TeV scale
quantum gravity and compact dimensions see \cite{shr}.

\section{Detection Rates}

The event-rate for an underground detector of area $A$ (events/sr/year) is calculable via

\begin{eqnarray}
{\mathrm Rate} = A \int_{E_{\mu}^{\mathrm{min}}}^{E^{\mathrm{max}}} dE_\nu\: P_\mu(E_\nu;E_\mu^{\rm min}) 
S(E_\nu){\frac{dN}{ dE_\nu}}. \label{rateqn}
\end{eqnarray}
 Here $P_\mu(E_\nu,E_\mu^{\rm min}$
is the  probability that a muon produced in a charged-current interaction 
arrives in a detector with an energy above the muon energy threshold 
$E_\mu^{\mathrm min}$ and  is given by 
\begin{equation} 
P_\mu(E_\nu,E_\mu^{\rm min}) = N_{\mathrm{A}}\, \sigma_{\rm CC}(E_\nu) \langle 
R(E_\nu;E_\mu^{\rm min} )\rangle , \label{pmudef} 
\end{equation} 
where 
$\langle R(E_\nu;E_\mu^{\rm min})\rangle$ is the average range of a 
muon in rock. $S(E_\nu)$ is the shadow factor defined earlier and 
$dN/dE_\nu$ is the (isotropic) neutrino flux.

 We give a set of sample rates below and refer the reader to   \cite{r98}
for a more extensive set of predictions..
 
Let us consider for illustration a detector with effective area 
$A=0.1\km^{2}$. This choice of size is intermediate to existing detectors
and the planned future facilities.  We show in Tables \ref{tab:eth1} and \ref{tab:eth10} 
the annual event rates for upward-going muons with observed energies 
exceeding $1\tev$ and $10\tev$, respectively.  We tabulate rates for 
the full upward-going solid angle of $2\pi$, as well as for the 
detection of ``nearly horizontal'' muons with nadir angle $\theta$ 
between $60^{\circ}$ and $90^{\circ}$.  The predicted event rates, 
shown here for the CTEQ4--DIS parton distributions, are very 
similar for other modern parton distributions.

We note that the atmospheric background overwhelms the signal when
the threshold is 1 TeV. However, signals should emerge above background 
for 10 TeV and 100 TeV thresholds. Above 100 TeV, we have the rare 
case of ``all signal and no background'', as the atmospheric muon
flux disappears. Also evident in Table 2 is the effect of shadowing,
with a majority of signal events in the ``nearly horizontal'' 
direction, where the neutrino traverses less rock prior to detection.

In addition to the (upward) partially contained events and the (downward)
fully contained ones which become important at energies above which the atmospheric background diappears , it may be possible to detect cascade neutrino interactions at extremely high energies ($\geq 10^{17}$ eV) in the proposed  Pierre Auger Cosmic Ray Observatory \cite{aug}. This  will consist of both a ground array  with 
detectors  distributed over a  very large  area ($\approx 3000$ km$^2$) and nitrogen fluorescence detectors. At these energies, the probability of a horizontally incoming neutrino interacting with the atmosphere is non-negligible. The events are thus rendered distinguishable from CR showers initiated by proton or other primaries by their incoming direction and their tendency to shower late into the atmosphere. 
The predictions for these   are  given in \cite{r98}.

\section{Conclusions}

 We have made an
attempt to review the  essential motivations and  phenomenological issues related to the 
interactions and detection of UHE neutrinos and provided expected event rates
for water/ice Cerenkov detectors.

Although present detector capabilities and sizes do not allow them to see above atmospheric backgrounds,
with upgradations and  several new  large-scale experiments underway, the detection of UHE neutrinos 
from astrophysical sources may be very likely in the near future.
 Besides 
the potential for serendipitous discovery, the unmatched (terrestially)
energies may lead to new particle physics discoveries. In addition, these 
experiments may help clarify the astrophysical mechanism responsible for 
the ultra-relativistic acceleration of matter in these sources and answer
important questions in CR physics, and, finally, provide sensitive tests 
of gravitational couplings and oscillations.
For those working in UHE neutrino physics and astronomy, the anticipation 
is palpable.

\section{Acknowledgements}

I thank Chris Quigg, Mary Hall Reno and Ina Sarcevic for  a stimulating 
collaboration,  several results of which are presented 
here.  I also  acknowledge useful discussions over the years  with
Pijush Bhattacharjee,  Gabor
Domokos, Susan-Kovesi Domokos and Francis Halzen, who have contributed to my knowlwdge of UHE neutrinos.





\begin{table}[htb]
	\caption{Upward $\mu^{+}+\mu^{-}$ event rates per year arising from 
	$\nu_{\mu}N$ and $\bar{\nu}_{\mu}N$ interactions in rock, 
for a detector with effective area $A = 0.1\km^{2}$ and muon energy 
threshol
d $E_\mu^{\mathrm{min}}=1\tev$.  The rates are shown 
integrated over all angles below the horizon and restricted to ``nearly 
horizontal''
 nadir angles $60^{\circ} < \theta < 90^\circ$.  }
\renewcommand{\arraystretch}{1.2}
	\begin{center}
\begin{tabular}{@{}lll}
  & \multicolumn{2}{c}{nadir angular acceptance} \\
Flux &  $0^\circ - 90^\circ$ &  $60^{\circ} - 90^\circ$  \\
\hline
 ATM \cite{volkova} &  1100. &  570.  \\
 ATM \cite{volkova} + charm \cite{prs} &  1100. &  570.  \\

 AGN-SS91 \cite{r56} & 500. & 380. \\
 AGN-M95 ($p\gamma$) \cite{r49} &  31.  &    18.    \\
 AGN-P96 ($p\gamma$) \cite{r48} &  45.  &   39.    \\ 
 GRB-WB \cite{waxb} &   12.  &    8.1  \\ 
 TD-SLSC \cite{slsc}  &   0.005 &    0.0046 \\ 
 TD-WMB12 \cite{r52} & 0.50 & 0.39 \\
 TD-WMB16 \cite{r52} & 0.00050 & 0.00039 \\
\end{tabular}
\end{center}
	\label{tab:eth1}
\end{table}

\begin{table}[htb]
\caption{Upward $\mu^{+}+\mu^{-}$ event rates per year arising from 
$\nu_{\mu}N$ and $\bar{\nu}_{\mu}N$ interactions in rock, for a 
detector with effective area $A = 0.1\km^{2}$ and muon energy 
threshold $E_\mu^{\mathrm{min}}=10\tev$.  The rates are shown 
integrated over all angles below the horizon and restricted to ``nearly 
horizontal'' nadir angles $60^{\circ} < \theta < 90^\circ$.}
\renewcommand{\arraystretch}{1.2}

	\begin{center}
\begin{tabular}{@{}lll}
  &  \multicolumn{2}{c}{nadir angular acceptance} \\
Flux &   $0^\circ - 90^\circ$ &   $60^{\circ} -  90^\circ$ \\
\hline
 ATM \cite{volkova} &  17. &   10. \\
 ATM \cite{volkova} + charm \cite{prs} & 19. & 11. \\
 AGN-SS91 \cite{r56} & 270. & 210. \\
 AGN-M95 ($p\gamma$) \cite{r49} &  5.7  &    4.3  \\
 AGN-P96 ($p\gamma$) \cite{r48} &   28.  &  25.  \\ 
 GRB-WB \cite{waxb} &    5.4  & 4.0 \\ 
\end{tabular}
\end{center}
	\label{tab:eth10}
\end{table}


\end{document}